\documentclass[letterpaper, 10 pt, conference]{ieeeconf}  

\IEEEoverridecommandlockouts                              

\usepackage{cite}
\usepackage{amsmath,amssymb,amsfonts}
\usepackage{algorithmic}
\usepackage{graphicx}
\usepackage{textcomp}
\usepackage{xcolor}
\usepackage{breqn}
\usepackage{subcaption}
\usepackage{float}

\title{\LARGE \bf
Condition Monitoring with Incomplete Data: An Integrated Variational Autoencoder and Distance Metric Framework}

\author{Maryam Ahang$^{1}$, Mostafa Abbasi$^{2}$, Todd Charter$^{1}$, and Homayoun Najjaran$^{1,2,*}$
\thanks{The authors are with the Advanced Control and Intelligent Systems Laboratory (ACIS),
        University of Victoria, Victoria, BC V8P 5C2, Canada
        {\tt\small maryamahang@uvic.ca; abbasi@uvic.ca toddch@uvic.ca; najjaran@uvic.ca}}%
\thanks{$^{1}$Department of Electrical and Computer Engineering,
        University of Victoria, Victoria, BC V8P 5C2, Canada}%
\thanks{$^{2}$Department of Mechanical Engineering,
        University of Victoria, Victoria, BC V8P 5C2, Canada}%
\thanks{$^{*}$Corresponding Author}
\thanks{\textit{ArXiv preprint}}
}

\usepackage{tikz}

\newcommand\copyrighttext{%
  \footnotesize \textcopyright \the\year{} IEEE. Personal use of this material is permitted. Permission from IEEE must be obtained for all other uses, including reprinting/republishing this material for advertising or promotional purposes, collecting new collected works for resale or redistribution to servers or lists, or reuse of any copyrighted component of this work in other works.}

\newcommand\copyrightnotice{%
\begin{tikzpicture}[remember picture,overlay]
\node[anchor=south,yshift=10pt] at (current page.south) {\fbox{\parbox{\dimexpr0.75\textwidth-\fboxsep-\fboxrule\relax}{\copyrighttext}}};
\end{tikzpicture}%
}

\begin{document}

\maketitle
\copyrightnotice
\thispagestyle{empty}
\pagestyle{empty}

\begin{abstract}
Condition monitoring of industrial systems is crucial for ensuring safety and maintenance planning, yet notable challenges arise in real-world settings due to the limited or non-existent availability of fault samples. This paper introduces an innovative solution to this problem by proposing a new method for fault detection and condition monitoring for unseen data. Adopting an approach inspired by zero-shot learning, our method can identify faults and assign a relative health index to various operational conditions. Typically, we have plenty of data on normal operations, some data on compromised conditions, and very few (if any) samples of severe faults. We use a variational autoencoder to capture the probabilistic distribution of previously seen and new unseen conditions. The health status is determined by comparing each sample's deviation from a normal operation reference distribution in the latent space. Faults are detected by establishing a threshold for the health indexes, allowing the model to identify severe, unseen faults with high accuracy, even amidst noise. We validate our approach using the run-to-failure IMS-bearing dataset and compare it with other methods. The health indexes generated by our model closely match the established descriptive model of bearing wear, attesting to the robustness and reliability of our method. These findings highlight the potential of our methodology in augmenting fault detection capabilities within industrial domains, thereby contributing to heightened safety protocols and optimized maintenance practices.

\end{abstract}


\section{Introduction}

Condition monitoring plays an inevitable role in the safety of any industrial system, especially when it comes to the sensitive parts of machines, like the bearings in rotating machinery, which are prone to faults. Machine learning (ML) and artificial intelligence (AI) techniques have gained a lot of attention in recent years in the field of condition monitoring \cite{liu2018artificial}. Despite their high accuracy, these methods usually rely heavily on data quality and availability. One of the main challenges in this field is not having complete or sufficient data \cite{ahang2024intelligent}. Systems usually operate in normal working conditions, so the normal data is ample. However, limited fault data can be collected, and such fault samples might not be representative of severe conditions in an actual plant, as regular maintenance usually addresses them. Some methods, like synthesizing fault samples and data augmentation, can generate more fault samples \cite{ahang2022synthesizing}. However, such methods can not be used for unseen or incomplete data, for example, when more severe faults happen for the first time.

Identifying the severity of faults in the industry is challenging for maintenance teams and data analysts who monitor the sensor data. Traditional data-driven condition monitoring methods often rely on the availability of ample labeled data for effective classification. However, the necessity for integrating unlabeled data is crucial, especially in instances where new categories are observed without any pre-existing examples. This scenario often necessitates the use of semi-supervised learning algorithms. These algorithms employ techniques to enhance label prediction accuracy and utilize regularization methods to prevent overfitting, ensuring reliable performance on unseen data \cite{Basu2009}.
In various fields like computer vision and large language models, different techniques such as domain adaptation \cite{berthelot2021adamatch}, transfer learning \cite{shi2009extending}, and ensemble classifiers \cite{de2021reliable} have been proposed to handle unseen, incomplete and unlabeled data. However, applying these methods in safety-critical domains like human health or industrial fault detection can be problematic due to the high cost and liability associated with real-life data. Additionally, these applications require a high level of robustness \cite{KHAN2021130}. Some classification methods, like one-class classification (OCC), have been utilized to detect new objects which do not fall into current classes \cite{HAYASHI2024119975}. This strategy involves training the model solely on normal segments of the target data, enabling early detection of deterioration. The model can provide early warnings of anomalies by applying OCC techniques, as they learn exclusively from normal conditions. However, OCC methods can only detect the anomalies, not their severity.
Constructing a unified semantic vector for condition monitoring based on theoretical correlations or distance functions applied to vibration data has been overlooked yet shows promising potential. Xu and Li pioneered this approach \cite{9670760}, employing a deep learning framework with a CNN-based feature extractor and generative adversarial module. They achieved a 77.03\% accuracy in identifying unseen faults by measuring the distance between extracted and generated features. Similarly, \cite{zarchi2023novel} introduced Weighted Neural Networks (WNN) to enhance diagnostic accuracy and feature transferability across domains, leveraging mean discrepancy functions. In \cite{CHEN2024110883}, an online unsupervised optimization model has been proposed, employing distance-metric, learning-based clustering to enhance separability between normal and degradation stages using high-dimensional features. However, a crucial missing component in fault detection is a deep learning-based distance function capable of detecting unseen classes, identifying the severity level and performing denoising operations by improving feature transferability.

\begin{figure*}[t]
  \centering
  \includegraphics[width=\linewidth]{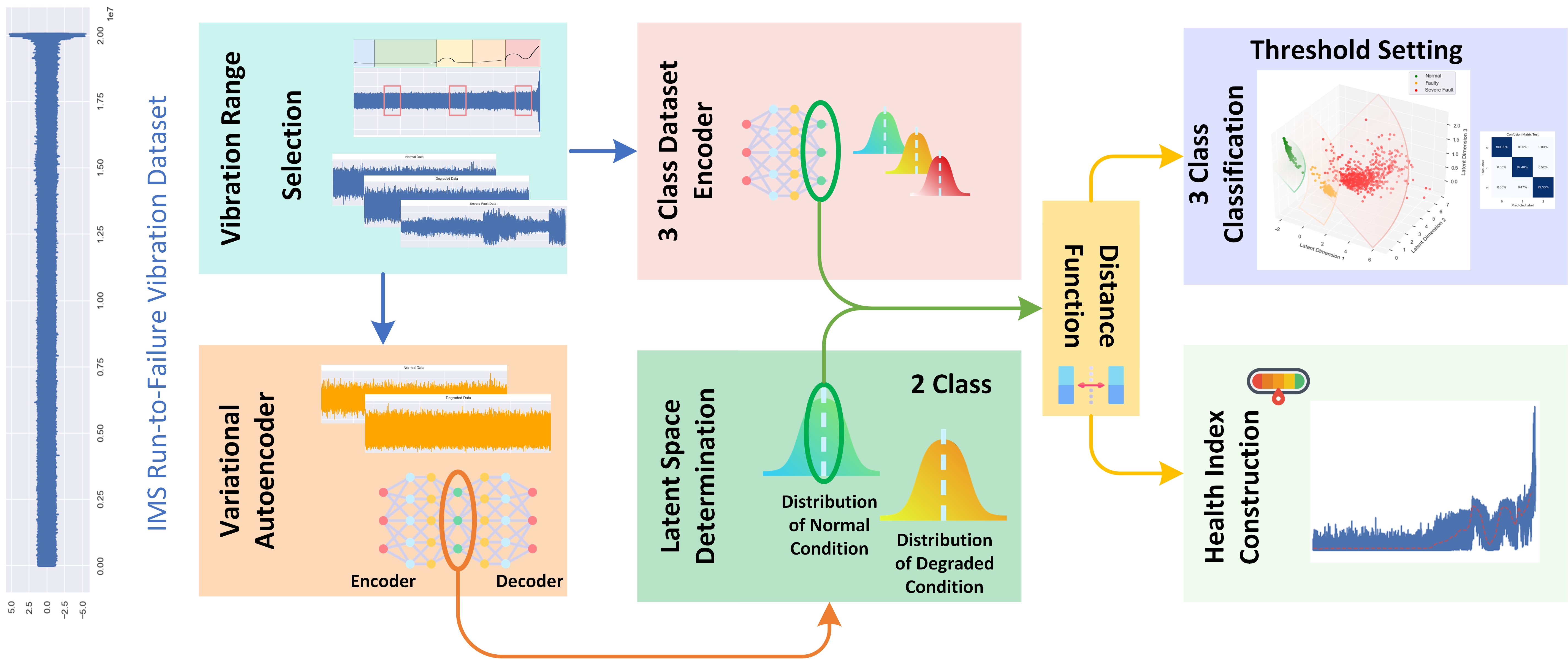}
  \caption{The framework of the proposed method$^{1}$}
  \label{frame}
\end{figure*}

This research introduces a novel method for handling incomplete data based on a Variational Autoencoder (VAE) and distance metrics. The data is considered incomplete if it does not contain all the information on the machine's useful life. In this paper, three different working conditions are considered for the system. Normal is the condition when the system is working in a normal operating state. Degraded is when the system condition deviates from the normal condition, but the system is still functional; it can be considered the starting point of developing faults. The final category, severe faulty condition, occurs when the system's condition deviates significantly from the normal due to a fault, leading to failure or drastically reduced performance. As the first step in the proposed methodology, a VAE is used for feature extraction and dimension reduction of the normal and degraded data (severe faults are considered unseen and unavailable data). This deep neural network maps the input data to probabilistic distributions in a latent space, where each class has a unique distribution. Then, the distance of each data point in the latent space from the mean of the normal data is calculated; the further a data point deviates from the normal condition, the worse the health state of the system becomes. This distance is used to generate a health index for the system. With the trained encoder part of the VAE, the mapping can be applied to any new data points, even for unseen conditions, and the corresponding health index can be calculated. The training phase of the proposed method is performed solely on normal and degraded data. Severe faults are introduced for testing, and a health index is assigned to the data showcasing the deterioration of the system over time. We apply the method using the IMS-bearing dataset as a benchmark for validation. The calculated health index closely follows the theoretical model of wear evolution introduced in \cite{el2014descriptive}. To validate the ability of the model to classify unseen faults, a classifier was designed based on thresholding the health indexes. The model could classify the new unseen faults with an accuracy of 99.51\%. A flow diagram of the proposed method and condition monitoring steps is shown in Fig.~\ref{frame}.

\section{Methodology}

\subsection{Variational Autoencoder}
Autoencoders are unsupervised learning algorithms that use backpropagation to produce an output that matches the input. They are made of two parts, an encoder and a decoder. The encoder maps input data to a lower dimension space, and the decoder reconstructs inputs from the latent space. So, the autoencoder can extract hidden relations between data, and we can use the low-dimension representation for further operations \cite{liu2018artificial}. Dimension reduction simplifies data by reducing the number of features used to describe it. This can be achieved by selecting a subset of the existing features or extracting new features that capture the essential information from the original dataset. After training an autoencoder, the latent space represents the principal features of the data, enabling further analysis, such as the classification of the new reduced feature space.

Stacked Autoencoders are multilayer neural networks that consist of one input and multiple hidden layers. They are similar to auto-associative neural networks (AANN), which are multilayer neural networks that make the input and output similar. The output of each hidden layer is used as the input of the next layer \cite{lu2017fault}.

A variational autoencoder is an autoencoder with regularized training to avoid overfitting and ensure that the latent space has good properties that enable a generative process. It maps the input to probabilistic distributions instead of a single point and can produce new data.
At first, the input is encoded as a probabilistic distribution in latent space, and then a point is sampled from that distribution. After decoding the sample, reconstruction error is computed and propagated through the network. The distributions generated by the decoder should closely resemble a standard normal distribution, ensuring that the latent space is properly regularized and facilitates the generation of new data.
The variational autoencoder loss function is composed of a reconstruction term, which makes data reproduction possible with a minimum error like a conventional autoencoder, and a regularization term, which makes latent space distributions similar to standard normal. This loss function is expressed as the Kulback-Leibler divergence between the returned distribution and a standard Gaussian. Improving the regularization term and span of the whole latent space increases reconstruction error, so a trade-off should be considered in that case.
The loss function of a variational autoencoder is described in \eqref{eq:vae}, where $x$ is the input to the VAE, $f(z)$ is the output of the decoder, $q_x (z)$ is Gaussian distribution approximating $p(z|x)$
considering $z$ as the latent variable, $p(z)$ is a standard Gaussian distribution, and $c$ is a constant balancing factor.

\begin{equation}\label{eq:vae}
\begin{split}
(f^*,g^*, h^*)= argmax(\mathbb{E}_{z\sim {q}_{z}}(-||x-f(z)||^2)/2c \\ - KL(q_x (z),p(z)))
\end{split}
\end{equation}

To improve the performance and robustness of the network, the coefficient $\beta$ is added to the base loss function of the autoencoder as shown in \eqref{eq:vae2}. A higher $\beta$ value increases the spread of data in the latent space, leading to better separation between data clusters. In a noisy environment, the noise will not have a major effect due to the distance between different groups of data, so the classification accuracy will be increased \cite{higgins2017beta}.

\begin{equation}\label{eq:vae2}
\begin{split}
(f^*,g^*, h^*)= argmax(\mathbb{E}_{z\sim {q}_{z}}(-||x-f(z)||^2)/2c - \\ \beta KL(q_x (z),p(z)))
\end{split}
\end{equation}

\subsection{Distance Metrics}

To establish a reliable health index for the system under different working conditions, we analyze the information from the latent space of the variational autoencoder. This approach leverages the VAE's dimension reduction and feature extraction capabilities. The distance between each data point and the mean of the normal data in the latent space is calculated and used as the system's health index. As the trained encoder of the VAE can map any data sample (even the data that has not been seen in the training) into a point in the latent space, the health index can be calculated for any data point.
Different distance metrics are calculated and compared, including Euclidean distance ($d_{E}$), Manhattan distance ($d_{T}$), also known as taxi-cab, and Minkowski distance ($d_{M}$) with the norm order of three, given by

\begin{equation}\label{eq:euclidean}
\begin{split}
  d_{E}(p,q)   = \sqrt {\sum _{i=1}^{n}  ( q_{i}-p_{i})^2 }
\end{split}
\end{equation}

\begin{equation}\label{eq:Manhattan}
\begin{split}
  d_{T}(p,q)   = {\sum _{i=1}^{n}  |q_{i}-p_{i}| }
\end{split}
\end{equation}

\begin{equation}\label{eq:Minkowski}
\begin{split}
  d_{M}(p,q)   = ({\sum _{i=1}^{n}  |q_{i}-p_{i}|^p })^ {\!1/p}
\end{split}
\end{equation}

\noindent respectively, where $p_{i}$ and $q_{i}$ are data points, $n$ is the number of dimensions, and $p$ represents the order of the norm.

\footnotetext[1]{This figure was prepared using images from www.flaticon.com.}

\section{Experiment and Analysis}

This section presents the experimental analysis of the study. It details the chosen case study, the characteristics of the dataset used for the experiment, and the network's specific architecture and implementation details. Finally, the condition monitoring results are presented and compared to those achieved by other methods, evaluating the effectiveness of the proposed approach.

\subsection{Dataset Description}

This research leverages the run-to-failure bearing dataset from the IMS centre \cite{qiu2006wavelet} as the case study. The setup consists of four Rexnord ZA-2115 double-row bearings on a shaft with a constant rotational speed of 2000 RPM by an AC motor and a radial load of 6000 lbs added to the shaft and bearings. The data is recorded from normal conditions until failure occurs after exceeding the designed lifetime. Data is collected with high-sensitivity quartz ICP accelerometers installed on the bearing's housing. Three data sets from a run-to-failure experiment are detailed, each with a sampling rate of 20 kHz, comprising 204,800 data points per sample. The faults observed in the bearings at the experiment's conclusion are as follows: In the first set, bearing 3 showed an inner race fault, and bearing 4 had a roller element defect. In the second set, bearing 1 exhibited an outer race defect. Lastly, in the third set, an outer race failure was noted in bearing 3.

The second set was frequently used as a benchmark \cite{li2019novel} and is the main data considered in this research.  The data contains 982 effective files. Since the dataset consists of run-to-failure data and is not labeled, we infer the health state of the bearings based on \cite{zhang2020semi}. For instance, in the second set, bearing 1 is considered to begin degrading after the 710\textsuperscript{th} data file. The data following this point is labeled as degraded. We then selected severe fault samples near the end of the dataset, just before the complete failure occurred. Normal data is chosen from near the beginning of the process, excluding the very beginning when the bearing is still in the running-in phase. Data is categorized into three labels: normal condition, degraded condition (minor faults), and severe faults. To train the VAE, a total of 20 data files were selected, representing both normal and degraded bearing conditions. Subsequently, these signals were segmented into discrete sub-samples, each encompassing 256 data points. 
Following segmentation, the samples were randomly shuffled to ensure unbiased selection. Finally, the data was split into training and testing sets at a 75\% and 25\% ratio, respectively. This prepared dataset was then used to train the VAE model. To evaluate the model's capability for unseen data, 10 data files from the severe fault condition were also added to the base dataset.

\subsection{VAE Implementation Details}

\begin{figure}[ht]
  \centering
  \includegraphics[width=.85\linewidth]{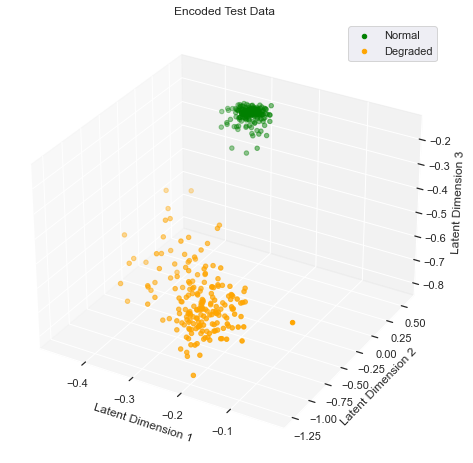}  
\caption{Three-dimensional representation of the latent space of the VAE using normal and degraded data}
\label{2DLatent_Train}
\end{figure}

The designed VAE consists of a four-layer encoder and decoder with relu activation functions. The encoder in this architecture is designed as a fully connected Multilayer Perceptron (MLP) with layer sizes of 256, 128, 32, and 8 neurons in sequence. Correspondingly, the decoder reverses this configuration, following the neuron sequence of 8, 32, 128, and 256. The latent space is set to be five-dimensional, meaning that each input vector is encoded to a five-dimensional probabilistic distribution. The training is done over 500 epochs, with Adam optimizer with a learning rate of 5e-4. The $\beta$ in the VAE's loss function is set to 20, and $c$ is considered 1. Latent encodings of the testing phases are visualized in Fig.\ref{2DLatent_Train}, showing three of the five dimensions. The VAE successfully maps all the normal and degraded data to their respective distributions, and these classes are easily separable from each other within this latent space.

All experiments were performed using Python 3.6 on a computer with an NVIDIA GeForce RTX 3090 GPU and an AMD Ryzen 9 processor with 64 GB of memory. 

\begin{figure}[htbp]
\centerline{\includegraphics[width=.85\linewidth]{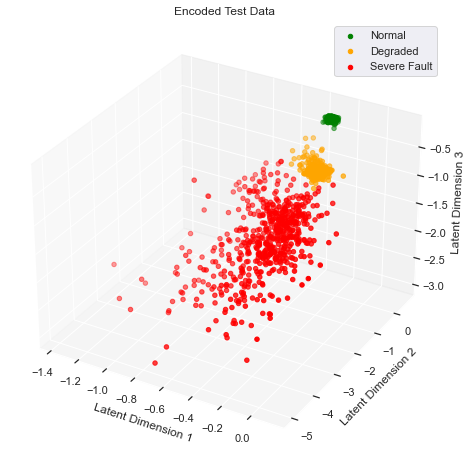}}
\caption{Latent space of the VAE depicting the seen and unseen classes}
\label{3class}
\end{figure}

To evaluate the effectiveness of the proposed method in detecting the new faults and unseen samples, we used the trained encoder of the VAE to map a new dataset containing the unseen faults (severe faults) to the latent space. The new latent space is visualized in Fig. \ref{3class}. It should be noted that the actual latent space is five-dimensional, and these plots only show three-dimensional representations. Yet, even with just three dimensions, the ability of the proposed method to separate different classes is evident.

\subsection{Health Index Assignment}

The efficiency of an optimized VAE in mapping the data in a rich latent space is shown in the previous subsection. The normal, degraded and very faulty data are separable, and the level of the degradation of the system is related to the distance from the normal samples. The worse the condition of the system, the further from normal data. This assumption allows us to assign a health index to every new and unseen data sample. We define the health index as the distance between each encoded data point from the mean of the encoded normal condition data in the five-dimensional latent space.

\begin{figure}[htbp]
\centering
\begin{subfigure}{0.45\textwidth}
  \centering
  \includegraphics[width=\linewidth]{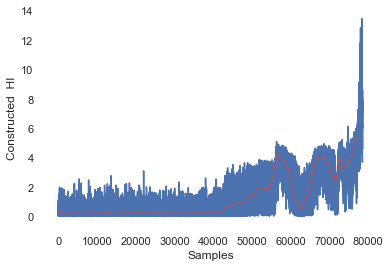}
  \caption{Health index of the bearing over time}
  \label{fig:HI}
\end{subfigure}
\hfill
\begin{subfigure}{0.45\textwidth}
  \centering
  \includegraphics[width=\linewidth]{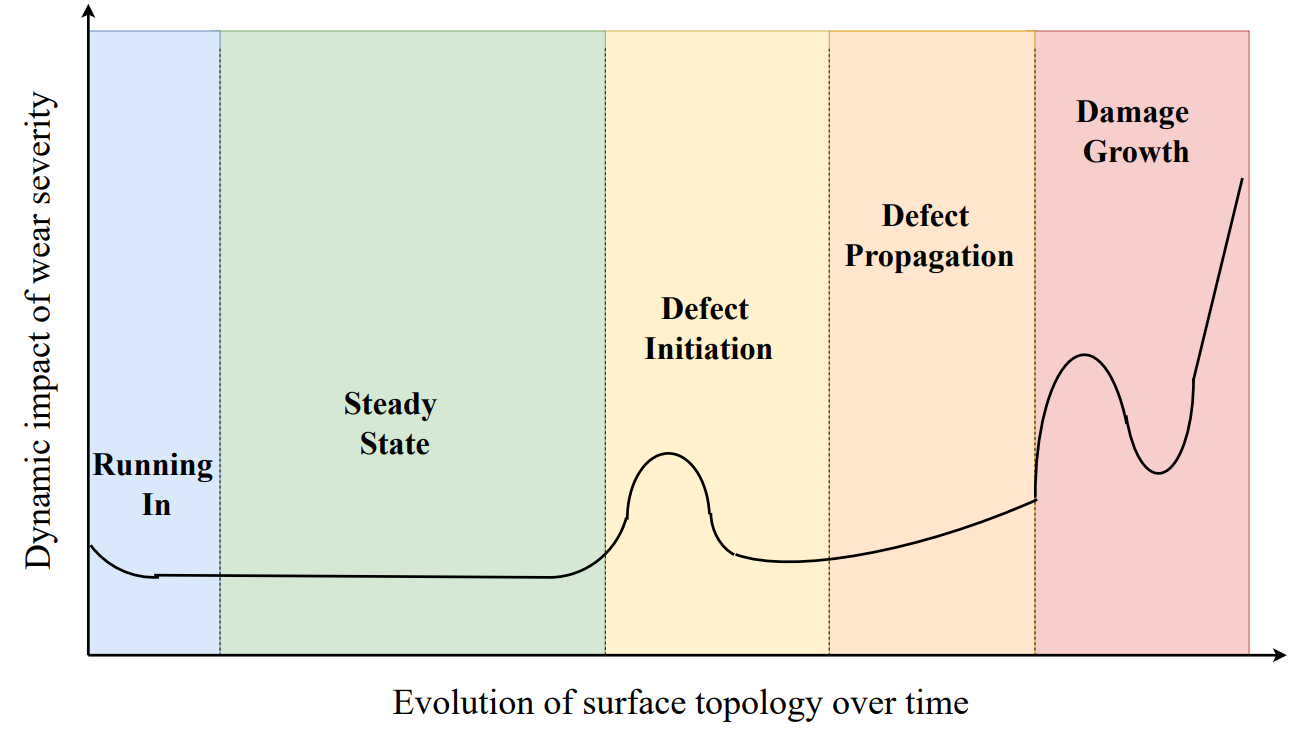}
  \caption{Dynamic behaviour of bearings due to wear evolution \cite{el2014descriptive}}
  \label{fig:HIR}
\end{subfigure}
\caption{Comparison of health index and wear evolution}
\label{fig:both}
\end{figure}

To validate the calculated health indexes, the health index for all the data points in the run-to-failure process, even for unseen data, is calculated and shown in Fig. \ref{fig:HI}. Which very closely follows the pattern of the dynamic impact of wear sensitivity described in Fig.  \ref{fig:HIR}\cite{el2014descriptive}. This finding suggests that the model can effectively track the different stages of wear progression. It is worth mentioning that for health index assignment, only a few samples of normal and degraded data are used to train the autoencoder, approximately 2\% of all data, yet the assigned health index is very accurate.

We employed data containing only normal and degraded samples to set thresholds on the assigned health indexes for these classes. The threshold for the normal class is the maximum distance between the mean of the normal data and each sample of the normal data in the dataset, and the threshold for the degraded data is the maximum distance of degraded samples set from the mean of normal data in the training. Subsequently, unseen severe faults were introduced in the test data, where after assigning the health index, fault detection is performed based on the previously defined thresholds. The model can detect the unseen classes with an accuracy of 99.51\%, and the overall detection accuracy is 99.84\%.

We evaluated the performance of three distance metrics (Euclidean, Manhattan, and Minkowski) for calculating the health index. The detailed comparison is presented in Table \ref{Dist}. While all three metrics showed promising results, Euclidean distance demonstrated a slight advantage over the others. Therefore, it has been selected as the health index in this study.

\begin{table}[htbp]
\caption{Comparison between different distance metrics}
\begin{center}
\label{Dist}
\begin{tabular}{lcccc}
\hline
\textbf{Distance Metric}& \textbf{Accuracy} & \textbf{Precision} & \textbf{Recall} & \textbf{F1 Score} \\
\hline
Euclidean        &  \textbf{0.9983}        &        \textbf{0.9983}  &  \textbf{0.9984}        &   \textbf{0.9983}  \\
Manhattan       &   0.9783       &    0.9791      &     0.9788      &    0.9784    \\ 
Minkowski    &   0.9866       &     0.9870      &     0.9866      &   0.9868    \\ \hline
\end{tabular}
\end{center}
\end{table}

To evaluate the generalizability of our proposed solution, we tested it on two additional subsets of the IMS dataset. The results, presented in Table \ref{Datasets}, are encouraging, suggesting the method's effectiveness.

\begin{table}[htbp]
\caption{Different bearings datasets}
\label{Datasets}
\begin{center}
\begin{tabular}{lllll}
\hline
\textbf{Distance Metric}& \textbf{Accuracy} & \textbf{Precision} & \textbf{Recall} & \textbf{F1 Score} \\
\hline
Subset 1 Bearing 3       &   0.9115       &    0.9231      &    0.9121       &   0.9134     \\
Subset 2 Bearing 1     &  0.9983        &        0.9983  &  0.9984       &   0.9983  \\ 
Subset 3 Bearing 3     &      0.9182    &   0.9346       &    0.9142       &   0.9151    \\ \hline
\end{tabular}
\end{center}
\end{table}

\begin{figure}[htbp]
\centerline{\includegraphics[width=.85\linewidth]{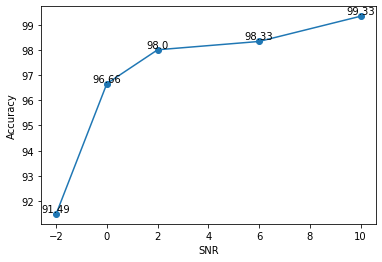}}
\caption{Analysis of robustness against noise}
\label{noise}
\end{figure}

To study the impact of noise on the proposed method, additive white Gaussian noise (AWGN) is added to the base signal. The signal-to-noise ratio (SNR) is the ratio of the power of the clean signal to the power of the noises. Five noise levels with an SNR between -2 and 10 dB are studied based on the noise level utilization in \cite{chen2021bearing}. The classification accuracy regarding the noise levels pictured in Fig. \ref{noise} shows the robustness and accuracy of the proposed model in noisy environments. 

\subsection{Comparative Analysis}

We conducted a comparative analysis to assess the performance of our proposed model against established benchmarks. We specifically compared our model with four prominent approaches: K-Nearest Neighbors (KNN), a Convolutional Neural Network trained for Zero-Shot Learning (ZSL-CNN), K-Means clustering, and a Vanilla Autoencoder. To ensure a fair comparison and leverage the strengths of our semantic construction approach, we integrate it into each model during the iterative training process. We then evaluated the test accuracy of all models to determine the efficacy of the proposed method.

Consistent with our proposed approach, all models aimed to classify data into three categories: normal, degraded, and severe fault. Notably, the introduction of a reliable normal reference was crucial for this classification. This method facilitated each model to distinguish degraded data from the reference normal data with the addition of a distance function and a threshold to detect severe, unseen faults.

\begin{figure}[ht]
    \centering
    \includegraphics[width=1\linewidth]{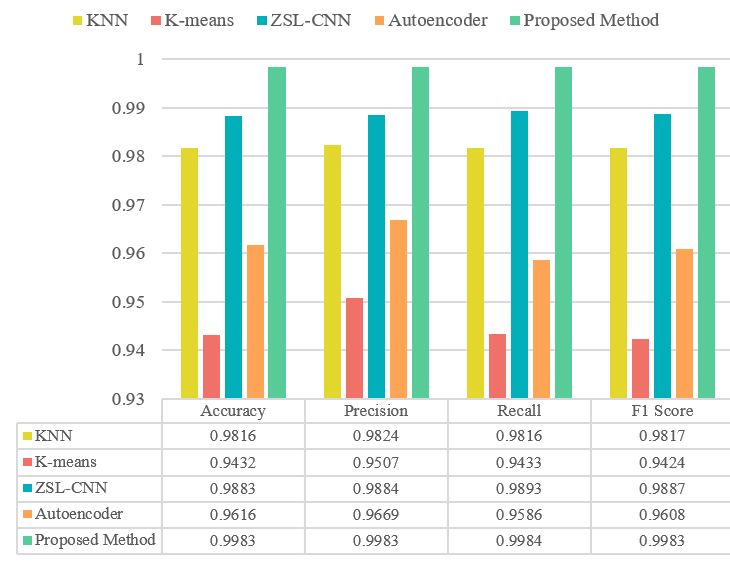}
    \caption{Comparison of evaluation metrics of different methods}
    \label{fig:comparison}
\end{figure}

The distance to the k-nearest neighbours for a data point is compared to a threshold established based on the maximum distance observed among degraded points. This comparison enables the identification of a third class, severe faults. Similarly, vector embeddings were employed in the CNN model to represent normal and degraded data references. Subsequently, the model quantifies a similarity between the embedding and data points, ultimately classifying unseen data into a new class based on the maximum distance. In the K-Means approach, centroids act as central reference points for each cluster. These centroids are used to compute the distance between new data points and their closest cluster. A vanilla autoencoder with the same structure as the proposed VAE is also utilized for comparison.
In Fig.~\ref{fig:comparison}, the test results of different models are presented for comparison.

The proposed method outperforms other approaches in terms of accuracy, precision, recall, and F1 score. While all methods achieve acceptable accuracy for unseen class detection, even a slight improvement in critical systems can significantly impact condition monitoring. As illustrated in the figure, it is evident our proposed method achieves superior performance, followed by ZSL-CNN, KNN, Autoencoder, and K-means.

\section{Conclusions and Future work}

In the context of industrial plant operations, the critical task of condition monitoring for safety and maintenance is often hindered by the scarcity or complete absence of fault samples in real-world scenarios. This research proposes a novel method for condition monitoring of industrial systems with incomplete data. Leveraging abundant normal and degraded data, the method utilizes a variational autoencoder for feature extraction and defines a distance-based health index on the VAE's latent space. This approach effectively tracks system health and detects unseen faults with 99.51\% accuracy. Additionally, the defined health indexes align with established wear evolution patterns. Our method demonstrates superior performance compared to existing unsupervised and semi-supervised approaches (KNN, KMeans, zero-shot learning CNN, Vanilla Autoencoder). Moreover, a robustness analysis shows that the proposed method performs well in the presence of noise, which is commonly prevalent in industrial environments. Future work will explore the potential of VAEs for sensor fusion in more complex systems and investigate more advanced thresholding algorithms.

\section*{Acknowledgment}
We would like to express our gratitude for the financial support provided by NTWIST Inc., FortisBC, and the Natural Sciences and Engineering Research Council (NSERC) Canada under the Alliance Grants ALLRP 555220 – 20 and ALLRP 557088 – 20. We also extend our appreciation for the collaboration of Fraunhofer IEM, Düspohl Gmbh, and Encoway Gmbh from Germany in this research.

\bibliographystyle{ieeetr}
\bibliography{bib}

\vspace{12pt}

\end{document}